\documentclass[fleqn,twoside]{article}
\usepackage{espcrc2}

\usepackage{graphicx}
\usepackage[figuresright]{rotating}

\newcommand{\bkappa}{\mbox{\boldmath $\kappa$}}

\newcommand{\bp}{\mbox{\boldmath $p$}}

\newcommand{\bb}{\mbox{\boldmath $b$}}
\newcommand{\br}{\mbox{\boldmath $r$}}

\def\lsim{\mathrel{\rlap{\lower4pt\hbox{\hskip1pt$\sim$}}
    \raise1pt\hbox{$<$}}}         

\newcommand{\AmS}{{\protect\the\textfont2
  A\kern-.1667em\lower.5ex\hbox{M}\kern-.125emS}}

\title{$k_\perp$--factorization for Hard Processes on Nuclei 
in the Saturation Regime}

\author{Wolfgang Sch\"afer  \address[IKP]{Institut f\"ur Kernphysik (Theorie), 
        Forschungszentrum J\"ulich, 
        D-52425 J\"ulich, Germany}} %

\begin{document}

\begin{abstract}
We discuss applications of the recently developed nonlinear
$k_\perp$--factorization for hard processes on nuclei. The starting
point is the color dipole approach which allows to introduce the
concept of a collective unintegrated nuclear gluon distribution.
We focus on single--jet spectra, which depend on the collective
nuclear glue in a nonlinear manner. We briefly comment on the Cronin effect.

\vspace{1pc}
\end{abstract}

\maketitle

\section{Dipole cross section $\leftrightarrow$ unintegrated glue}

Deep inelastic scattering at very small $x \ll 1$ is conveniently 
viewed as a scattering of multiparton Fock states 
of the virtual photon,
which propagate along straight--line, fixed--impact parameter trajectories 
and interact coherently with the target. The proper formalisation 
valid for inclusive, as well as
diffractive deep inelastic processes results in the color--dipole
approach \cite{Nikolaev:1990ja}. 
The total virtual photoabsorption cross section takes the
well-known quantum mechanical form $\sigma_{tot}(\gamma^*p;x,Q^2)
= \langle \gamma^*| \sigma(\br) |\gamma^*\rangle$ and becomes 
calculable in terms of the $q\bar{q}$--lightcone-wavefunction 
of the virtual photon and the color dipole--nucleon cross 
section $\sigma(\br)$. The connection between
color--dipole formulas and $k_\perp$--factorization is
provided by
$\sigma(\br) =  \sigma_0 \, \int
d^2{\mbox{\boldmath$\kappa$}} [1 -
e^{i{\mbox{\boldmath$\kappa$}}{\br}}]
f({\mbox{\boldmath$\kappa$}})$, where
$f({\mbox{\boldmath$\kappa$}})$ is directly related to the
unintegrated gluon distribution $ f({\mbox{\boldmath$\kappa$}}) =
{4 \pi \alpha_S \over N_c \sigma_0}\,  {1\over
{\mbox{\boldmath$\kappa$}}^4} \, \partial
G_N/\partial\log({\mbox{\boldmath$\kappa$}}^2)$. 
In DIS off free nucleons, at HERA energies, the $s$--channel
partial waves are small, $\sigma_N(\mathrm{diffractive})/\sigma_N(\mathrm{total}) \sim
0.1$. The constraints from unitarity do not matter, parton
densities are a meaningful concept, and the 
linear $k_\perp$--factorization is appropriate. There is 
a drastic change when going to heavy nuclear targets. 
Now the unitarity constraints
become saturated, and at small $x$ one expects
$\sigma_A(\mathrm{diffractive})/\sigma_A(\mathrm{total}) \to
0.5$ \cite{KolyaTalk}. The $q \bar{q}$--Fock state of the virtual 
photon is coherent over the whole nucleus 
for $x \lsim x_A =1/m_N R_A \ll 1$, where $m_N$ is the nucleon mass, 
and $R_A$ the nuclear radius. Glauber--Gribov
theory prescribes how to construct the $s$--channel unitarized 
dipole--nucleus scattering amplitude $\Gamma_A[\bb,\sigma({\br})] = 1-\exp[-\sigma({\br}) T(\bb)/2]$, where $T(\bb)$ is the nuclear thickness as a function of 
the impact parameter $\bb$ which labels the partial waves. 
They define the collective nuclear unintegrated glue $\phi(\bb,x_A,\bkappa)$:
\begin{equation} 
\Gamma_A[\bb,\sigma(\br)] = \int d^2\bkappa\phi(\bb,x_A,\bkappa)[1-\exp(i\bkappa\br)]   
\end{equation}
A salient feature of the strongly absorptive regime is that 
multiple scattering/unitarization introduces the saturation scale
$Q_A^2  \approx 4  \pi^2/N_c \cdot \alpha_S(Q_A^2) G(x_A, Q_A^2) T(\bb)$.
The collective nuclear unintegrated gluon density $\phi(\bb,x_A,\bkappa)$ behaves
like its free--nucleon counterpart in a number of interesting applications:
both the total DIS cross section  as well as diffractive DIS amplitudes exhibit the linear 
$k_\perp$--factorization property. For these observables the nuclear dependence
is fully contained in $\phi(\bb,x_A,\bkappa)$, which simply takes 
the place of the free--nucleon quantity $f(\bkappa)$ in the relevant formulas.
Due to the color--singlet nature of the virtual photon, the 
same holds true for single--jet 
cross sections in the photon fragmentation region. 
For the case of fully diferential dijet cross sections in DIS
we observed however that the linear $k_\perp$--factorization is
broken \cite{Nikolaev:2003zf}. 
Still, the collective nuclear glue $\phi$ remains a useful quantity,
and a new, nonlinear $k_\perp$--factorization emerges. For a further discussion
of DIS including the nonlinear small--$x$ evolution and references, see \cite{KolyaTalk}. 

\section{Inclusive Production $\equiv$ Excitation of Fock States of the Beam
Parton}

Here we report on recent progress \cite{SingleJet} in extending the nonlinear nuclear 
$k_\perp$--factorization to processes of the type $a^* \to bc$,
such as excitation of open charm and jets, $g^* \to Q\bar{Q},
g^* \to g_1 g_2, q^* \to qg$ in hadron--nucleon/nucleus scattering and/or DIS 
at $x \ll x_A$. From the lab frame point of view all of the above processes
can be viewed as excitation of the perturbative $ |bc\rangle$ Fock state 
of the physical parton $|a\rangle_{phys} = |a\rangle_0 +
\Psi(\br) |bc\rangle_0$, by means of multiple gluon exchanges 
with the target nucleus. Here $\Psi(\br)$ is the light--cone wave--function
for the transition $a \to bc$, with $\br$ the transverse distance between $b$ and $c$
which are moving at impact parameters $\bb_b,\bb_c$.  
\begin{figure}[htb]
 \vspace{9pt}
 \includegraphics[scale=0.28,angle=270]{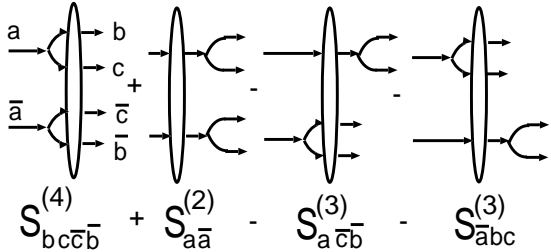}
 \caption{Diagrammatic representation for the evolution operator of the four--particle density matrix.}
 \label{fig:DensityMatrix}
 \end{figure}
All partons $a,b,c$ are supposed to have 
rapidities $\eta_{a,b,c} > \eta_A = \log1/x_A$, so that our considerations are 
relevant e.g. for the region of large (pseudo)rapidities in $pA (dA)$ 
collisions at RHIC. 
In terms of the bare parton $S$--matrices, 
the excitation amplitude takes the simple form
\begin{equation}
\hat{f}(a\to bc) = \Big[S_b(\bb_b)S_c(\bb_c) -S_a(\bb_a)\Big]
\Psi({\br}) |bc\rangle
\end{equation}
After squaring, there
emerges a master formula for the final state density matrix whose structure
is indicated in Fig. \ref{fig:DensityMatrix}.  It involves 
$S^{(4,3,2)}$ which are appropriate matrix elements of the intranuclear
evolution operator for a four(three,two)--particle system, coupled
to an overall color--singlet state, what ensures the necessary
color gauge cancellations in an elegant way. We
stress that the intranuclear evolution operator is a matrix in the
space of singlet four-parton dipole states $|R{\bar{R}}\rangle
=|(bc)_R \otimes (\bar{b}\bar{c})_{\bar{R}}\rangle$, further
details depend on the color representations of the partons
involved. For the single--jet spectra, the transverse momentum of $c$
is integrated over, and the problem becomes a single--channel 
one. When applying the formalism to the free nucleon target in the 
single--gluon exchange approximation, one obtains the standard results 
of linear $k_\perp$--factorization, with the gain of 
very convenient representations in terms of light--cone wavefunctions. 
For the nuclear target, all the pertinent single--jet cross section
have unambiguous representations in terms of the collective nuclear glue 
$\phi(\bb,x_A,\bkappa)$, but now they become manifestly \emph{nonlinear}
functionals of the target glue. For illustration we cite the single--particle
spectrum for $g^* \to Q\bar{Q}$:
\begin{eqnarray}
\lefteqn{ {(2\pi)^2 d\sigma_A(g^* \to Q\bar{Q}) \over dz d^2\bp d^2\bb}= 
 S_{abs}(\bb) \int d^2\bkappa \phi(\bb,x_A,\bkappa)}
\nonumber \\
\lefteqn{ \Big\{|\Psi(z,\bp) - \Psi(z,\bp+z\bkappa) |^2 +} 
\nonumber \\
\lefteqn{ 
|\Psi(z,\bp +\bkappa) -
\Psi(z,\bp+z\bkappa)|^2\Big\} }
\nonumber \\
\lefteqn{ + \int d^2\bkappa_1 d^2\bkappa_2 \phi(\bb,x_A,\bkappa_1) \phi(\bb,x_A,\bkappa_2)}
\nonumber \\
\lefteqn{
|\Psi(z,\bp+\bkappa_2) -\Psi(z,\bp+z\bkappa_1+z\bkappa_2) |^2 \, .}
\label{eq:5.5}
\end{eqnarray}
Here $z$ is the fraction of the lightcone--momentum of the gluon carried 
by the measured jet. The first term is linear in $\phi$, has the same structure as 
the free nucleon cross section, and is suppressed by the absorption factor
$ S_{abs}(\bb) = \exp[-\sigma_0 T(\bb)/2]$, which vanishes in the strong 
absorption limit ($\sigma_0$,
the dipole cross section for large dipoles). The dominant contribution at 
strong absorption is the second one, which is quadratic in $\phi$.
Similar formulas have been 
obtained for the transitions $q^* \to q(\bp)g$ as well as $q^*\to q g(\bp)$,
which both show a quadratic nonlinearity, and for $g^* \to g_1 g_2$, which
displays a cubic dependence on the nuclear glue \cite{SingleJet}. Here we point out that 
also the relevant contribution to integrated inclusive cross sections
will exhibit a nonlinear dependence on $\phi$.
\begin{figure}[htb]
 \vspace{9pt}
 \includegraphics[scale=0.33]{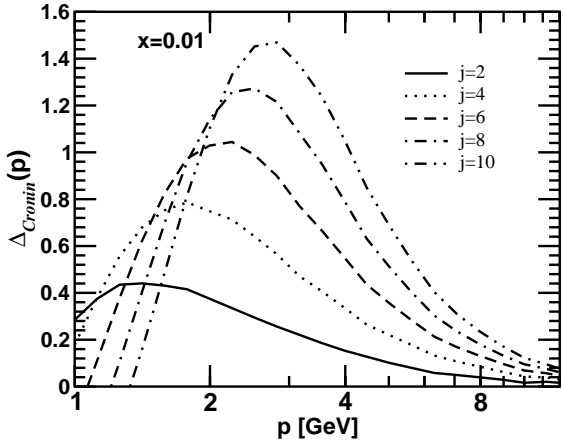}
 \caption{Parton level Cronin enhancement shown for different effective numbers 
of rescatterings, $j$.}
 \label{fig:Cronin}
 \end{figure}
Two simple corollaries on the excitation of  \emph{slow}, $z \ll 1$ 
partons are worth to mention: First, for all processes cited above,
the spectra of \emph{slow} partons universally allow a linear 
$k_\perp$--factorization. Second, the spectrum of \emph{slow} gluons in DIS from 
the excitation $\gamma^* \to q \bar{q} g$ also allows for an
exact linear $k_\perp$--factorization representation of the form
\begin{eqnarray}
{z_g d\sigma_A^{(\gamma^* \to q\bar{q}g)} \over dz_g d^2\bp}
\propto {1 \over \bp^2} \int d^2\bkappa {\partial G_A(\bkappa) \over \partial \bkappa^2}
{\partial G_{\gamma^*}(\bp-\bkappa) \over \partial(\bp- \bkappa)^2}\, .
\nonumber
\end{eqnarray}
These findings are in line with the results obtained on slow gluon production
in \cite{KovchegovMueller}.
For practical applications the linear factorization of slow parton 
spectra is of limited relevance though: one would have to ask
for the transverse momentum $\bp$ of the produced jet to be 
compensated by particles from a widely separated region of rapidity.
Our results being fully differential in both transverse momentum $\bp$ as well as 
longitudinal momentum parametrised by $z$ also solve the problem
of the $\bp$--dependence of the Landau--Pomeranchuk--Migdal effect
for the limit of a thin target (see also \cite{SlavaTalk}).
Finally we comment on the relevance of nonlinear $k_\perp$--factorization
to the Cronin effect, the enhancement of high--$\bp$ particle production
at intermediate, but still hard, $|\bp|$ in hadron--nucleus collisions.
If one decomposes the single jet cross section into the contributions
of the direct and resolved interactions of the fragmenting parton $a$,
then the Cronin effect can be shown to be a nuclear modification of the
resolved part \cite{SingleJet}. The seed for the Cronin effect is contained in the 
antishadowing enhancement of the collective nuclear glue first discussed in
\cite{Nikolaev:2000sh}. Remarkably, although the saturation scale for a realistic
nuclear glue is a modest $Q_A \sim 1$ GeV, the Cronin enhancement takes place
at perturbatively large $|\bp| \sim 2 \div 2.5$ GeV. In Fig.\ref{fig:Cronin}
we show the Cronin effect in the resolved parton interactions as a function
of the transverse momentum $\bp$ of the jet parton for several multiplicities 
of effective rescatterings $j$. The production of slow quarks off a nucleus
of mass number $A\sim 200$ corresponds to $j=4$, for slow gluons read $j=8$. 
The effect for observed hadrons will be placed at smaller $|\bp|$, and 
the contribution from direct interactions shall dilute the effect, the
corresponding evaluations are in progress. 

{\small
I should like to thank the organizers of Diffraction 2004 for 
invitation to a most pleasant and informative workshop.}

\end{document}